\documentclass[12pt]{article}
\usepackage{a4wide}
\usepackage{amssymb}
\begin{document}
\newcommand{\case}[2]{{\textstyle \frac{#1}{#2}}}
\newcommand{\lP}{l_{\mathrm P}}

\newcommand{\md}{{\mathrm{d}}}

{\pagestyle{empty}
{\renewcommand{\thefootnote}{\fnsymbol{footnote}}
\begin{center}
\mbox{} \hfill CGPG--03/5--2
\vfill 
{\LARGE Initial Conditions for a Universe\footnote{This
essay was awarded First Prize in the Gravity Research Foundation Essay 
Contest 2003.}}\\
\vfill 
Martin Bojowald\footnote{e-mail
address: {\tt bojowald@gravity.psu.edu}} \\
\vspace{0.5em}
Center for Gravitational Physics and Geometry,\\
The Pennsylvania State
University,\\
104 Davey Lab, University Park, PA 16802, USA\\
\vspace{1.5em}
\end{center}
}

\setcounter{footnote}{0}

\vfill

\begin{abstract}
 In physical theories, boundary or initial conditions play the role of
 selecting special situations which can be described by a theory with
 its general laws. Cosmology has long been suspected to be different
 in that its fundamental theory should explain the fact that we can
 observe only one particular realization. This is not realized,
 however, in the classical formulation and in its conventional
 quantization; the situation is even worse due to the singularity
 problem. In recent years, a new formulation of quantum cosmology has
 been developed which is based on quantum geometry, a candidate for a
 theory of quantum gravity. Here, the dynamical law and initial
 conditions turn out to be linked intimately, in combination with a
 solution of the singularity problem.
\end{abstract}

\vfill\vfill

\mbox{}


\mbox{}

\newpage
}
\setcounter{page}{1}
{

By design, physical theories provide a framework to deal with a large
class of situations in such a way that a variety of different
phenomena can be seen to have their origin in a small number of basic
physical concepts. As an example, Maxwell theory links seemingly
unrelated observations in optics and electricity as properties of the
electromagnetic field. Usually, a theory also contains rules how to
specify boundary or initial conditions in order to select a special
class of systems within a vast range of possibilities which can be
realized, e.g., in a certain experimental setup. The particular choice
of those conditions, however, is left open by the theory.

In cosmology, the theory of the universe as a whole, the situation has
long been expected to be different: as observers, we have access only
to one particular realization of the universe, and its initial
conditions cannot be changed. This should be reflected in the
fundamental theory of the cosmos; initial conditions for a universe
should be part of the theory, rather than the choice of a theorist.

In classical cosmology as described by general relativity, however,
the situation is different, even worse thanks to the singularity
problem according to which a universe like our own has to start with a
``big bang'' singularity. At such a point the theory breaks down and
initial conditions cannot be imposed there.

To illustrate this, we can look at the simplest case which is an
isotropic universe with flat space. Its dynamical law, derived from
Einstein's field equations, is the Friedmann equation
\begin{equation} \label{Friedmann}
 \left(\frac{\dot{a}}{a}\right)^2=\case{16\pi}{3}G\rho(a)
\end{equation}
which describes the evolution of the radius $a(t)$ of the universe as
a function of time. If we know the gravitational constant $G$ and the
matter content which enters via the energy density $\rho(a)$, we can
determine the evolution. For a particular realization, we have to choose
initial values $a(t_0)$ and values of possible matter fields at some
initial time $t_0$.

Ideally, $t_0$ would be the ``creation time'' of the universe where
the initial conditions are either chosen or, hopefully, predicted by
the theory. However, in classical cosmology the initial time
represents a singularity where the theory breaks down. For instance,
if we choose the matter content to be pure radiation the energy
density $\rho(a)$ is proportional to $a^{-4}$ and any solution of the
Friedmann equation has the form $a(t)\propto\sqrt{t-t_0}$. At $t_0$
the radius of the universe vanishes which implies that energy
densities or tidal forces are infinite and the evolution as described
by the Friedmann equation breaks down. Those are the unmistakable
signs of a singularity, which can be reached in a finite amount of an
observer's time but presents a boundary to what the theory can tell
us. There is no way to tell what happens beyond the singularity or if
there even is any ``beyond the singularity''.

The cosmological singularity is often viewed as the point of creation
of the cosmos via a big bang. But initial conditions cannot be imposed
there since the evolution equation (\ref{Friedmann}) would give us an
infinite time derivative of $a$. Instead, we have to choose another
time where the system is not singular and impose initial conditions
there which then are completely arbitrary.

The singularity presents a problem by itself which is often hoped to
be cured by quantization. In fact there is reason to be optimistic
since also in quantum mechanics a classical problem, the instability
of the Hydrogen atom, is solved by the presence of a finite ground
state energy $E_0=-\frac{1}{2}m_ee^4/\hbar^2$. Up to inessential
constant factors this is the only non-relativistic energy value which
can be obtained from fundamental constants just for dimensional
reasons. Without $\hbar$, there would simply not be any natural value
for a possible lowest energy. Moreover, one can see that it is
important to quantize because $\hbar$ appears in the
denominator. Thus, in the classical limit $\hbar\to0$ the ground state
energy diverges and we return to instability. As we know, there are
also other effects of $E_0$, most importantly the discreteness of the
energy spectrum.

For gravity, we can use a similar argument: Its fundamental parameter
is the gravitational constant $G$ which, together with $\hbar$ gives
us a natural length, the Planck length $\lP=\sqrt{G\hbar}\approx
10^{-33}\mbox{cm}$. If the Hydrogen atom is any indication, we can
expect to have a smallest length in a quantum theory of gravity which
would lead to a very different behavior close to the cosmological
singularity. We do not notice this length scale in experiments because
it is so tiny, but it should have important implications in physical
situations which involve small scales. In the classical limit
$\hbar\to0$, the minimal length would approach zero and we can expect
to see the singularity problem arise in this way. Furthermore, we
can also anticipate that the presence of $\lP$ implies a discreteness
of space or length spectra. The explicit form of such a spectrum can
only be derived from a detailed theory, but its presence can be
expected purely on dimensional grounds.

Thus, it seems worthwhile to quantize cosmology; but it is not
expected to be straightforward: since classical cosmology is part of
general relativity, we need at least a part of a quantum theory of
gravity. An approach tailored to simple models as the one we discussed
before, is the Wheeler--DeWitt quantization. We replace $\dot{a}$ in
the Friedmann equation by the momentum $p_a=3a\dot{a}/8\pi G$ and use
the familiar quantum mechanical procedure to replace $p_a$ with the
operator $\hat{p}_a=-i\hbar\md/\md a$ acting on a wave function
$\psi(a)$. Choosing an ordering of operators, we obtain the
Wheeler--DeWitt equation
\begin{equation} \label{WdW}
 -\frac{1}{6}\lP^4a^{-1}\frac{\md}{\md a}a^{-1}\frac{\md}{\md a}a\psi=
  8\pi G\hat{H}(a)\psi
\end{equation}
where $\hat{H}(a)$ is a matter Hamiltonian which we do not need to
specify for our purposes. This equation is our dynamical law,
presenting an evolution equation in the ``internal time'' $a$ which
means that the evolution of possible matter fields is measured with
respect to the expansion or contraction of the universe. Concerning
the singularity problem there is no real progress because the equation
cannot tell us about anything happening beyond the singularity at
$a=0$.

The issue of initial conditions now appears in a different light: we
have to choose an initial value for the wave function $\psi(a)$ at
some $a_0$ (we only need one value to fix one of the two parameters of
the general solution; the other one would be fixed by
normalization). At $a=0$, corresponding to the classical singularity,
the differential equation is still singular, but we can try to cancel
the divergence by requiring the initial condition $\psi(0)=0$. This is
DeWitt's initial condition \cite{DeWitt}, and it seems that we do have
a relation between this initial condition and the dynamical law since
it was motivated by a regularity condition. Unfortunately, this is not
true since the uniqueness of this condition depends on the
matter content as well as the factor ordering. Even worse, however, is
the fact that in a more complicated system DeWitt's initial condition
would not be well-posed: the only solution satisfying it would vanish
identically.

There are attempts \cite{SIC} to make DeWitt's condition well-posed in
general by adding a ``Planck potential'' to the Wheeler--DeWitt
equation solely for the purpose of creating one (and only one)
solution which decreases toward zero at $a=0$ such that it can be
hand-picked by the initial condition. This presents another attempt to
link the dynamical law with the initial condition, but the
introduction of the Planck potential and the choice of the wave
function remain artificial.

Later, DeWitt's condition has been replaced by alternative proposals
which originate from different motivations, most importantly the
tunneling proposal of Vilenkin's \cite{tunneling} and the no-boundary
proposal of Hartle and Hawking's \cite{nobound}. They are not directly
related to the dynamical law, however, and they do not solve the
singularity problem.

Was the hope originating in the stability of the quantized Hydrogen
atom misleading? Do we have to accept the cosmological singularity and
the fact that we still have to choose our initial conditions even for
a whole universe? Maybe surprisingly, the answer is not a certain
Yes. For we have used only simple quantum mechanics to derive the
quantum model, while a full quantum theory of gravity in this spirit
exists only formally and a precise link is lacking. The full theory
would be much more complicated and it would have to fulfill many
consistency conditions which can easily be missed in a simple model
with only a single gravitational degree of freedom, $a$. There is in
fact one indication that the quantization we used is not correct:
while the Planck length $\lP$ does appear in the Wheeler--DeWitt
equation (\ref{WdW}), there is no realization of discreteness of space
as we would have expected ($a$ can still take arbitrary continuous
values).

The situation has changed over the last decade since we now have a
mathematically well-defined candidate for quantum gravity (loop
quantum gravity/quantum geometry \cite{Rov:Loops,ThomasRev}) from
which we can derive quantum cosmological models (see \cite{IsoCosmo}
and references therein). There are many consistency conditions to
fulfill which leads to a theory very different from the
Wheeler--DeWitt quantization; in particular, they imply that space is
in fact discrete. For our model we need the following information: The
wave function $\psi_n$ is now only defined at integer values $n$
related to $a$ by $a_n^2=\frac{1}{6}\lP^2|n|$ rather than on a
continuous line, and the total volume of space can only take discrete
values
\begin{equation}\label{Vol}
 V_n=(\case{1}{6}\lP^2)^{\frac{3}{2}}
 \sqrt{(|n|-1)|n|(|n|+1)}\,.
\end{equation}
Here the Planck length appears and sets the scale for the discreteness
and the smallest non-zero volume $V_{\rm min}=\frac{1}{6}\lP^3$.

Over the last few years, the cosmological sector of quantum geometry,
loop quantum cosmology, has been studied. A first observation is that
we have a well-defined, {\em finite} operator which quantizes the
classically divergent $a^{-1}$ \cite{InvScale}. This operator has
eigenvalues
\begin{equation}
 (a^{-1})_n=16\lP^{-4}\left(\sqrt{V_{n+1}}-\sqrt{V_{n-1}}\right)^2
\end{equation}
in terms of (\ref{Vol}), which have the upper bound (for $n=2$)
\begin{equation}
 (a^{-1})_{\rm max}=\frac{32(2-\sqrt{2})}{3\lP}\,.
\end{equation}
Now we can see that also the second indication we got from the
Hydrogen atom is realized: The classical divergence of $a^{-1}$ is cut
off by quantum effects leading to an upper bound, which diverges in
the classical limit $\lP\to0$ owing to the appearance of the Planck
length in the denominator. Another surprising result is that
$(a^{-1})_0=0$, i.e.\ the inverse radius of the universe vanishes at
the classical singularity $n=0$ where also the radius itself
vanishes. This classically counterintuitive but well-understood fact
will be of importance later.

Thus, both facets of the Hydrogen atom are also present in our new
quantum cosmological model. To finally settle the singularity issue,
however, we still have to face the acid test: whether or not we can
extend the evolution to something ``beyond the singularity''. For this
we need the dynamical law, the loop quantization of the Friedmann
equation. It turns out to be \cite{Sing}
\begin{equation} \label{disc}
(V_{n+2}-V_n)
 \psi_{n+1}- 2(V_{n+1}-V_{n-1})
 \psi_{n}+(V_n-V_{n-2}) \psi_{n-1}\nonumber\\
= -\case{1}{3}8\pi G \lP^2\, \hat{H}(n)\,\psi_n
\end{equation}
where we use the volume eigenvalues (\ref{Vol}) and a matter
Hamiltonian $\hat{H}(n)$. It looks very different from the
Wheeler--DeWitt equation (\ref{WdW}), most obviously because it is a
difference rather than a differential equation. This is a direct
consequence of the discreteness of space and also time, which is now
given by the label $n$ instead of the continuous $a$. Nevertheless, it
is straightforward to check, using a Taylor expansion, that the
Wheeler--DeWitt equation approximates our discrete equation at large
volume $n\gg1$. When the volume is small, however, there are large
discrepancies between the discrete and the continuous formulation
which lead to qualitative changes. This is right where a modified
evolution is needed since we have seen that the Wheeler--DeWitt
formulation cannot deal with the singularity problem.

To check for a singularity we try to follow the evolution as long as
possible, starting with initial values for $\psi_n$ at two times $n_0$
and $n_0-1$ and evolving backward toward the classical singularity at
$n=0$. This is possible as long as the lowest coefficient,
$V_n-V_{n-2}$ in the difference equation is non-zero. At first one
can anticipate a problem because this coefficient is zero if (and only
if) $n=1$ such that we are not able to compute $\psi_0$, the wave
function at the classical singularity, in this way. This time,
however, we are safe: While we cannot find this value, we do not even
need it since it decouples completely from the evolution equation. Let
us ignore this value and try to continue the evolution, computing
$\psi_{-1}$ using (\ref{disc}) with $n=0$. Now two terms containing
the unknown $\psi_0$ appear, but both of them drop out. The first one,
is $2(V_{n+1}-V_{n-1}) \psi_{n}$, which is zero for $n=0$. We also
have $\case{1}{3}8\pi G \lP^2\, \hat{H}(n)\,\psi_n$ being zero for
$n=0$, but more subtly so: each term in the matter Hamiltonian, the
kinetic and the potential term, contains either components of the
metric or the inverse metric, reducing to $a$ or $a^{-1}$ in the
isotropic context. Classically, one would be zero and the other
infinite at the classical singularity, but we have seen that in loop
quantum cosmology {\em both} have to be zero at the classical
singularity. Thus, $\hat{H}(0)=0$ and $\psi_0$ completely drops out of
the evolution equation; $\psi_{-1}$ is completely determined by
$\psi_1$ which we know in terms of our initial data. In the same way,
we can now follow the evolution completely determining all values of
the wave function for positive and negative $n$. The evolution does
not stop at $n=0$ which, consequently, does not represent a
singularity anymore.

In this analysis the point $n=0$ was special because some coefficients
of the difference equation vanish. However, it does not represent a
singularity or a ``beginning'' of the universe. Instead, we can
determine what happens at the other side, represented by negative $n$,
by using our evolution equation. Intuitively, there is a collapsing
branch of the universe at negative times $n<0$ which collapses down to
a single point, bounces and enters our expanding
branch. Furthermore, one can show that the sign of $n$ is the
orientation of space such that the universe ``turns its inside out''
at $n=0$.

Without $n=0$ representing a beginning, it is not so natural to impose
initial conditions there; and it is not even possible because $\psi_0$
drops out of the evolution equation. Still, this point plays an
important role for the issue of initial conditions \cite{DynIn}, the
main interest of this essay. Let us take a closer look at what we
discussed before: Starting with initial values at $n_0$ and $n_0-1$ we
evolved backward until we reached $n=0$ and continued beyond the
singularity. At $n=0$ we noticed that we could not determine $\psi_0$,
which we just ignored because $\psi_0$ decoupled completely. However,
the part of the evolution equation which was supposed to give us
$\psi_0$ --- with $n=1$ in (\ref{disc}) --- still has to be satisfied,
resulting in a linear equation for $\psi_1$ and $\psi_2$. Those two
values, in turn, are linear functions of our two initial values
$\psi_{n_0}$ and $\psi_{n_0-1}$. Therefore, the dynamical law gives us
one linear condition for the two initial values, which is just what we
need to fix the wave function uniquely up to its norm.

Thus, for the first time we can now see an intimate link between the
dynamical law and initial conditions as part of the law. The discrete
structure, the solution of the singularity problem and the issue of
initial conditions are all related in a way which is very special to
the case of gravity and cosmology, for it is the Planck length which
allows the discreteness of space and the classical singularity problem
which makes the point $n=0$ special.

\medskip

The author is grateful to A.~Ashtekar and J.~Baez for discussions
which helped improve the interpretation of the results described here.
This work was supported in part by NSF grant PHY00-90091 and the
Eberly research funds of Penn State.

}


\begin{thebibliography}{10}

\bibitem{DeWitt}
B.~S.\ DeWitt,
\newblock Quantum Theory of Gravity. I. The Canonical Theory,
\newblock {\em Phys.\ Rev.} 160 (1967) 1113--1148

\bibitem{SIC}
H.~D.\ Conradi and H.~D.\ Zeh,
\newblock Quantum cosmology as an initial value problem,
\newblock {\em Phys.\ Lett.\ A} 154 (1991) 321--326

\bibitem{tunneling}
A.\ Vilenkin,
\newblock Quantum creation of universes,
\newblock {\em Phys.\ Rev.\ D} 30 (1984) 509--511

\bibitem{nobound}
J.~B.\ Hartle and S.~W.\ Hawking,
\newblock Wave function of the Universe,
\newblock {\em Phys.\ Rev.\ D} 28 (1983) 2960--2975

\bibitem{Rov:Loops}
C.\ Rovelli,
\newblock Loop Quantum Gravity,
\newblock {\em Living Reviews in Relativity} 1 (1998)
  http://www.livingreviews.org/Articles/Volume1/1998--1rovelli, [gr-qc/9710008]

\bibitem{ThomasRev}
T.\ Thiemann,
\newblock Introduction to Modern Canonical Quantum General Relativity,
\newblock {\em Liv.\ Rev.\ Rel.}, [gr-qc/0110034]

\bibitem{IsoCosmo}
M.\ Bojowald,
\newblock Isotropic Loop Quantum Cosmology,
\newblock {\em Class.\ Quantum Grav.} 19 (2002) 2717--2741, [gr-qc/0202077]

\bibitem{InvScale}
M.\ Bojowald,
\newblock Inverse Scale Factor in Isotropic Quantum Geometry,
\newblock {\em Phys.\ Rev.\ D} 64 (2001) 084018, [gr-qc/0105067]

\bibitem{Sing}
M.\ Bojowald,
\newblock Absence of a Singularity in Loop Quantum Cosmology,
\newblock {\em Phys.\ Rev.\ Lett.} 86 (2001) 5227--5230, [gr-qc/0102069]

\bibitem{DynIn}
M.\ Bojowald,
\newblock Dynamical Initial Conditions in Quantum Cosmology,
\newblock {\em Phys.\ Rev.\ Lett.} 87 (2001) 121301, [gr-qc/0104072]

\end{thebibliography}

\end{document}